# High-Performance Labyrinth Circular Bragg Grating Design for Charge and Stark-Tunable Quantum Light Sources Spanning Visible to Telecom Wavelengths

Rohit Prasad,[1] Quirin Buchinger,[1] Fei Chi Kristy Yuen,[1,2] , Yorick Reum,[1] Sven Höfling,[1] Tobias Huber-Loyola[1,3]

[1] Julius-Maximilians-Universität Würzburg, Physikalisches Institut, Lehrstuhl für Technische Physik, 97074 Würzburg, Germany

[2] University of British Columbia, Department of Physics and Astronomy, Vancouver, British Columbia, Canada, V6T 1Z1

[3] Karlsruhe Institute of Technology, Institute of Photonics and Quantum Electronics (IPQ) and Center for Integrated Quantum Science and Technology (IQST), 76131 Karlsruhe, Germany

## Abstract:

Semiconductor quantum dots embedded in circular Bragg gratings (CBGs) are among the most efficient integrated single-photon sources. However, the fully etched rings of conventional CBGs restrict the implementation of charge and Stark tuning via electrical contacts. To overcome this limitation, a labyrinth CBG geometry with four bridges has been proposed, yet the added bridges significantly degraded optical performance. In this work, we numerically demonstrate that a periodic labyrinth CBG design preserves both high coupling efficiency and strong Purcell enhancement while enabling electrical integration if optimized after introducing the bridges. We show three optimized designs at emission wavelengths of 780 nm, 930 nm, and 1550 nm, because these wavelengths are among the most relevant for quantum dots and show the general applicability of our approach. At all three wavelengths collection efficiencies exceeding 90% into a numerical aperture of 0.7 and Purcell factors greater than 25 are achieved. Furthermore, we propose a device layout incorporating a barrier layer that separates p- and n-doped semiconductor regions, which is incorporated to prevent tunneling of one of the charge carriers for selective charging. Also this design can be reoptimized to retain the performance of a device without tunnel barrier. These results establish labyrinth CBGs as a platform for electrically tunable quantum dot single-photon sources with high efficiency and scalability.

# Introduction

The generation of entangled photons is a critical requirement for the practical implementation of optical quantum computing [1] and to secure quantum communication networks [2]. With the growing adoption of one-way quantum computing architectures [3] and quantum cryptography protocols [4], there is an urgent demand for reliable entangled photon sources that simultaneously offer high efficiency, high indistinguishability, and spectral tunability. Among the various solid-state platforms [5-11] under investigation, semiconductor quantum dots [12] embedded in wavelength-scale optical cavities [13,14] have emerged as one of the most promising systems for generating on-demand entangled photons across visible[15], near-infrared [16,17], and telecom bands [18,19].

Achieving optimal performance from quantum dot emitters requires careful design of the photonic environment to enhance spontaneous emission via the Purcell effect [20-21] and to efficiently direct emitted photons into a well-defined optical mode [22-26]. For this purpose, micropillar cavities [27,28] and circular Bragg gratings (CBGs) [29-30] are two of the most prominent and widely studied nanophotonic cavity architectures. Each design offers distinct advantages: micropillars provide relatively straightforward electrical contacting [31-33], high quality factor confinement, high Purcell factors, and extraction efficiencies [34,35], while CBGs offer high Purcell factor and high extraction efficiencies in a broader spectral range [36-42] at the cost of a very demanding QD-CBG alignment[43,44].

However, a well-known challenge associated with standard CBGs is that the full angular etching of the cavity layers necessary to form the concentric grating pattern interrupts the conductive pathway needed for electrical injection, Stark tuning or deterministic charging of the quantum dot. In contrast a micropillar cavity provides a direct simpler contact point for the Stark tunability making it a more desirable cavity design [45,46].

To address this drawback in CBG design while retaining the desirable broadband operation and high Purcell factor, innovative design modifications have been proposed. One approach involves introducing narrow bridges creating a labyrinth design that connect the inner cavity region to the surrounding contacts, providing a pathway for electrical access without significantly disturbing the optical mode profile [47]. Another design strategy replaces the continuous grating with a pattern of concentric circular air holes [46,49] etched into the semiconductor making a hole-type CBG, where the unetched regions between the holes naturally form conductive pathways.

While the hole-type CBG has demonstrated excellent theoretical performance, with predicted collection efficiencies exceeding 90% for numerical apertures (NA) of 0.7, it requires considerable fabrication effort due to the complexity of defining and etching densely packed nanoscale hole patterns with high precision and uniformity. In contrast the labyrinth CBG design, which employs a limited number of narrow conductive bridges distributed symmetrically around the cavity, is far easier to fabricate using established lithographic and etching techniques. However, the presence of these bridges alters the optical properties of the optimized cavity, thereby reducing the overall efficiency of the CBG [48-52].

In this paper, we present a comprehensive theoretical study demonstrating that a four bridge labyrinth CBG design can achieve collection efficiency comparable to standard CBGs and the more complex hole-type CBGs across a broad wavelength range, while offering significant advantages in fabrication practicality. By simultaneously achieving high collection efficiency, robust cavity performance, and straightforward electrical contacting, the proposed design holds great promise as an efficient, electrically tunable source of entangled photons for future quantum photonic technologies. This work aims to motivate further experimental realization and integration of labyrinth CBG designs, with high-performance quantum light sources that are compatible with charge control and wavelength tuning via piezoelectric or electrical means.

## Device Design and Simulation Method

The proposed device uses a three-layer structure, a semiconductor layer containing the quantum dot at its center, a gold mirror below it to reflect downward-emitted light back into the upper hemisphere, and a transparent spacer that sets the distance to the mirror. The devices are designed for target wavelengths of 780 nm, 930 nm, and 1550 nm, each using materials suited for their spectral range. For the 780 nm design, aluminum gallium arsenide ($Al_{0.33}Ga_{0.67}As$, *n* = 3.35) serves as the semiconductor with an aluminum oxide ($Al_2O_3$, *n* = 1.627) spacer. For the 930 nm design, gallium arsenide (GaAs, *n* = 3.52) [53] is used with aluminum dioxide ($Al_2O_3$, *n* = 1.62) spacer. And for the 1550 nm design, indium aluminum gallium arsenide ($In_{0.53}Al_{0.23}Ga_{0.24}As$, *n* = 3.3) [54] serves as the semiconductor with an aluminum oxide ($Al_2O_3$, *n* = 1.617) spacer.

In addition, a 930 nm device incorporating an extra $Al_{0.8}Ga_{0.2}As$ (*n* = 3.01) [50] layer beneath the quantum dot layer, which is in the optical center of the semiconductor layer is investigated. This extra layer serves as a blocking barrier for the charges moving towards either the p- or n-doped GaAs region. The blocking barrier allows confinement of predominantly one charge carrier by blocking tunneling of one charge type. This additional simulation aims to examine whether the structure can be optimized to maintain collection efficiency and Purcell factor comparable to the design without the blocking barrier layer.

To create second-order Bragg reflection and strong in-plane confinement, concentric rings are etched into the semiconductor layer to form the CBG. These rings are connected using four narrow bridges establishing an electrical connection between the inner disc and the outer region, as presented by Buchinger, Q. et al. [45,50] and depicted in Figure 1a. The etched ring pattern starts with a defined inner radius and uses a constant grating period and etch width. For inner radius $R$, grating period $P$, and etch width $W$, the $n^{th}$ ring's inner and outer radii are calculated as $R + (n-1)P \pm \frac{W}{2}$. The bridge width was fixed at 125 nm to ensure a reliable electrical connection to the embedded quantum dots while minimizing the impact on the optical properties of the CBG. To ensure that the trenches are etched all the way down, even with moderate sidewall angles, the etch width is constrained to be at least 65 nm.

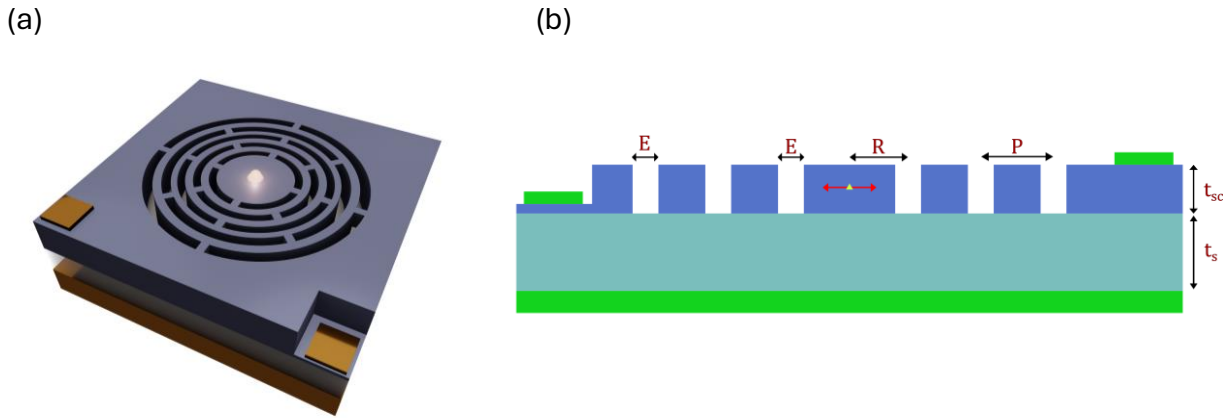

Figure 1: (a) 3D model of the labyrinth CBG design, showing the semiconductor membrane with gold contacts for Stark tuning, spaces and gold mirror for preferentially upwards emission of the photons. (b) Cross-section schematic of the structure, with R as the inner radius, E as the etch width, P as the grating period, $t_{sc}$ as the semiconductor thickness, and $t_s$ as the spacer thickness. The cut through the structure is chosen such that the bridges are avoided.

The optical performance of the full 3D cavity structures is simulated using Ansys Lumerical and FlexCompute Tidy3D commercial Finite-Difference Time-Domain (FDTD) solvers. The FDTD method is a numerical technique for solving Maxwell's equations suitable to simulate the propagation of electromagnetic waves in complex structures. The simulations are performed with cryogenic temperature refractive indices to match typical experimental operating conditions for semiconductor quantum dot emission. A non-uniform meshing scheme is applied, with the maximum grid size constrained to at least 100 mesh points per wavelength in the overall simulation area to ensure accurate resolution of the etched features while maintaining reasonable computational runtime.

To evaluate the collection efficiency, a box monitor enclosing the entire structure is employed. The efficiency into the upward hemisphere is calculated by taking the ratio of the power transmitted through the top monitors to the total power passing through the enclosing box. The collection efficiency into a given NA is obtained by calculating the far-field radiation pattern and integrating it within the corresponding collection cone and multiplying this fraction by the upper hemisphere efficiency. This approach provides an accurate estimate of the realistic collection efficiency that can be achieved using a high-NA objective in an experimental setup.

## Optimization and Constraints

Figure 1 illustrates the five key parameters that define our cavity design: inner radius R, grating period P, etch width E, semiconductor thickness $t_{sc}$, and spacer thickness $t_s$. The devices are optimized with the goal of maximizing collection efficiency within a numerical aperture (NA) of 0.7 while keeping the Purcell factor above 25. In a CBG structure, in-plane optical confinement is primarily controlled by the inner radius, the grating period of the concentric circles, and the etch width. The vertical confinement is determined by the semiconductor layer thickness, while the distance to the gold back mirror governs the interference between directly upward-emitted field and the downward-emitted reflected field. This configuration allows us to decouple the optimization of the in-plane parameters inner radius (R), grating period (P), and etch width (E) from the vertical parameters semiconductor thickness ($t_{sc}$) and spacer thickness ($t_s$).

The optimization of the device design is guided by a composite figure of merit (FOM), that balances five critical performance metrics: emission wavelength, Purcell factor, upper hemisphere efficiency, far-field Gaussian, and confined Gaussian emission within a specified numerical aperture (NA). The cost function to be minimized is defined as

$$Cost \ = \ w_1 \cdot f_1(\lambda) \ + w_2 \cdot f_2(P) \ - w_3 \cdot C_{NA=0.7}.$$

where λ is the peak emission wavelength, $P$ is the Purcell factor, $C_{NA=0.7}$ is the collected power through a mode-matched numerical aperture of 0.7. The weights $w_1, w_2$, and $w_3$ are user-defined and set to 0.2, 0.2, and 0.6, respectively. The individual components of the cost function are defined as $f_1(\lambda) = (\lambda - \lambda_0)^2$, this term penalizes deviation from the target resonant wavelength $\lambda_0$, $f_2(P) = (P - P_0)^2 \, (1 \ + \ e^{P - P_0 - 5})^{-1}$, where $P_0$ is the lowest Purcell factor acceptable for the optimization, set to 25 in this work. The sigmoid-like denominator ensures that penalties are relaxed for values of $P$ significantly exceeding $P_0$, thereby shifting the optimization focus to the other components of the cost function once an acceptable Purcell factor is achieved. And $C_{NA=0.7} \ = \ E_{up} \cdot \mathcal{O}_{ff} \cdot NA_{0.7}$, where

$E_{up}$ is the upward-emitted power, $\mathcal{O}_{ff}$ quantifies the Gaussian mode overlap of the farfield ,evaluated at 1m distance, with am arbitary Gaussian mode, and $NA_{0.7}$ denotes the fraction of the mode confined within a NA of 0.7. This formulation allows simultaneous optimization of wavelength targeting, radiative enhancement, and collection efficiency within realistic experimental constraints.

To efficiently optimize the device performance, the process is carried out in two stages. In the first stage, a simplified model consisting of a suspended semiconductor layer without the spacer or metallic (gold) layers. This simplification significantly reduces computational cost and simulation time. A Particle Swarm Optimization (PSO) algorithm is then employed to determine the optimal structural parameters.

In the second stage, the spacer and gold layers are incorporated to more accurately represent the final device structure. Using the PSO-optimized parameters as initial guesses, we perform a gradient-based optimization employing algorithms such as COBYLA and Powell to further refine the design. This hierarchical optimization approach yields the final optimized set of parameters, including the R, P, E, $t_{sc}$, $t_s$ while substantially reducing the overall optimization time and computational demand.

# Result

## Optimal Design and Parameter analysis

The optimal design parameters for the three devices, each tailored to operate at 780 nm, 930 nm, and 1550 nm, respectively, are summarized in Table 1. These parameters, including inner radius, grating period, etch width, semiconductor thickness, and spacer thickness, have been optimized to achieve Purcell factors of 27, 30, and 27, and corresponding collection efficiencies of 90.2%, 91.9%, and 91.2% within an NA of 0.7, respectively.

Table 1. Optimal design parameters for 780 nm, 930 nm, and 1550 nm devices

| Wavelength | | **780 nm** | **930 nm** | **1550 nm** |
|---|---|---|---|---|
| Inner radius | (nm) | 362 | 405 | 725 |
| Grating Period | (nm) | 313 | 357 | 649 |
| Etch width | (nm) | 68 | 80 | 146 |
| Semiconductor thickness | (nm) | 124 | 153 | 258 |
| Spacer thickness | (nm) | 172 | 200 | 328 |
| Purcell factor | | 27 | 30 | 27 |
| Collection efficiency NA 0.7 | | 90.2% | 91.9% | 91.2% |

Figures 2a–c illustrates the simulated Purcell factor and collection efficiency as a function of wavelength for each design, evaluated across various NAs. All three designs demonstrate collection efficiencies exceeding 90% within NA = 0.7 at and near their respective resonance wavelengths. Figures 2d–f further illustrates the collection efficiency at the resonance wavelength as a function of NA for the 780 nm, 930 nm, and 1550 nm designs, respectively. In radiative cascades, the polarization of emitted photons is intrinsically linked to their propagation direction, with orthogonal polarizations typically modeled as orthogonal dipole emitters. The degree of polarization entanglement between photon pairs critically depends on the spatial overlap of the two orthogonal far-field emission patterns [55]. Notably, incoherently summing up the far-field patterns from the orthogonal dipoles obscures this essential information. Therefore, the analysis was performed on the far-field emission pattern of a single dipole, as shown in the insets of Figures 2d–f.

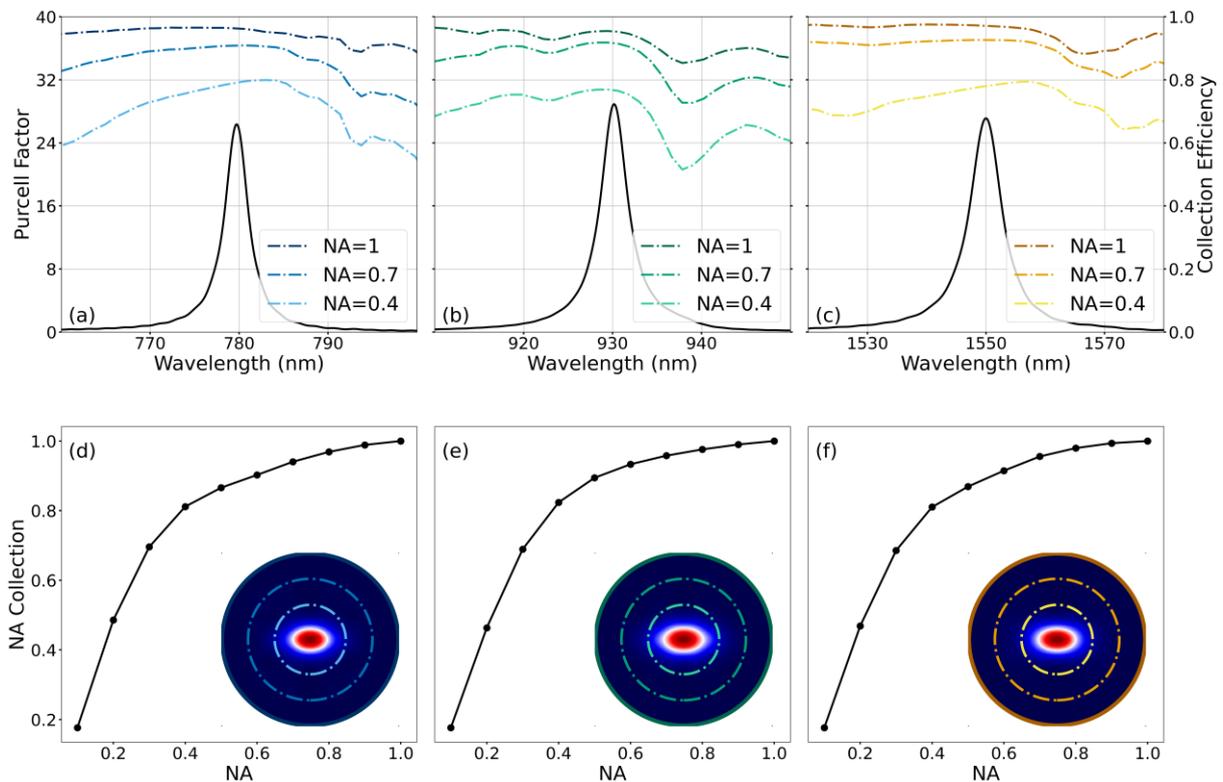

Figure 2. Purcell and collection efficiency of optimal design. (a)-(c) illustrates the 25+ Purcell and 90%+ $NA_{0.7}$ collection efficiency achieved by the optimal design for 780nm, 930nm, and 1550nm respectively. And (d)-(f) showcases the collection efficiency dependency on NA at the respective resonant wavelength. The inset shows the confined normalized far-field pattern from a single dipole. The indicated circles correspond to the NA from (a-c) 1, 0.7 and 0.4 respectively.

While the optimal parameters for these cavity designs can be determined relatively easily through numerical simulations, translating these theoretical designs into physical structures with nanometer scale precision remains a substantial experimental challenge. Figure 3 highlights the sensitivity of the system by illustrating the effects of 10 nm shifts in three key in-plane design parameters, the R, P, E. As well as a 5 nm shift in the $t_{sc}$ for each of the target wavelengths.

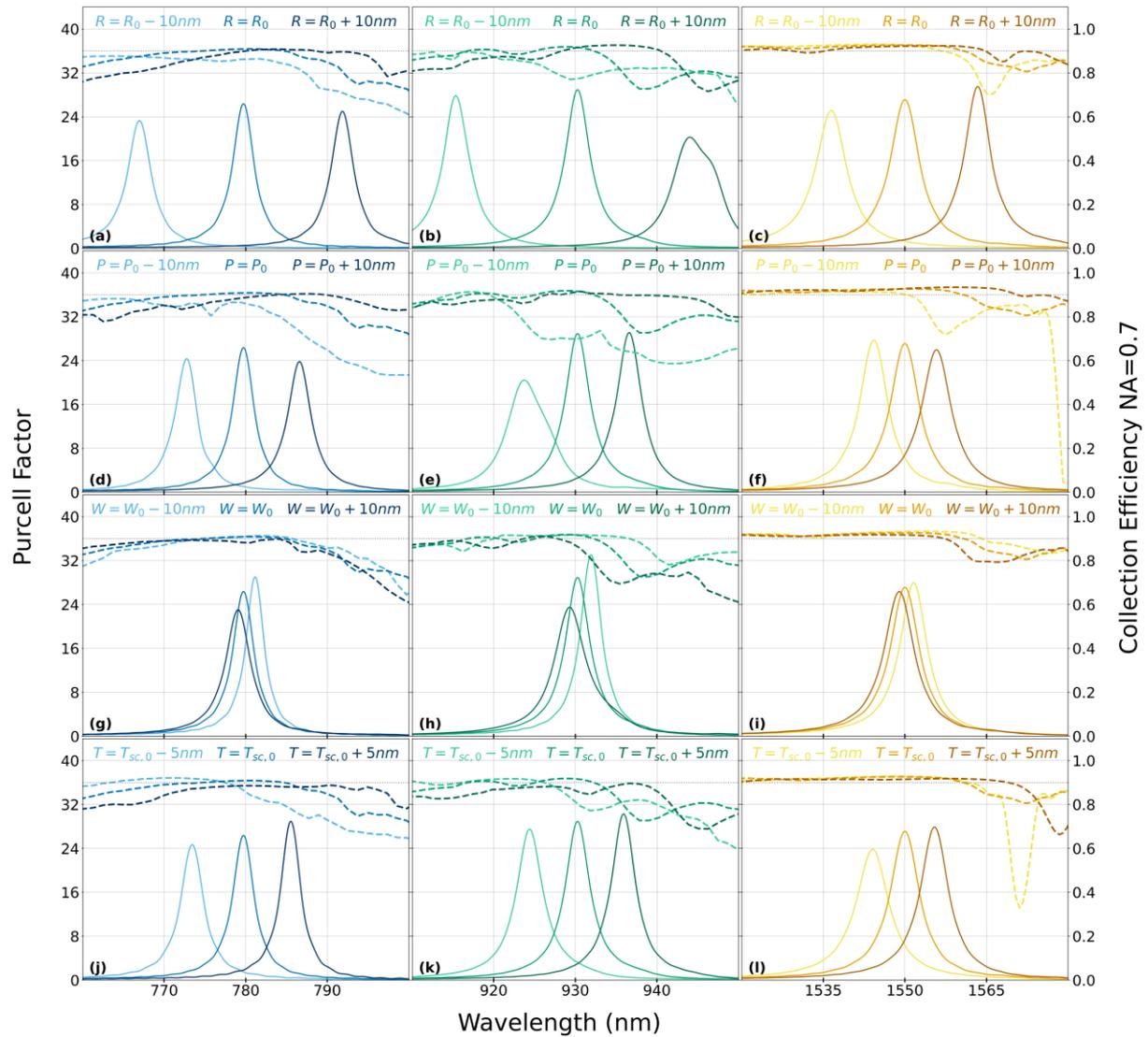

Figure 3. Effect of fabrication imperfections the Bragg parameters on device performance. Here $R_0, P_0, E_0, t_{sc,0}$ represent the optimized inner radius, Period, Etch width and semiconductor thickness. (a)-(c) Illustrates the impact of deviation of $\pm 10nm$ in the inner radius($R_0$), (d)-(f) Illustrates the impact of deviation of $\pm 10nm$ in the grating period($P_0$), (g)-(i) Illustrates the impact of deviation of $\pm 10nm$ in the etch width($E_0$), and (j)-(l) Illustrates the impact of deviation of $\pm 5nm$ in the semiconductor thickness($t_{sc,0}$) on the resonant wavelength and collection efficiency within NA = 0.7 for 780nm, 930nm, and 1550nm design, respectively.

These parameters play an essential role in defining the optical confinement of the mode and, as a result, strongly influence the resonance condition of the cavity. Figure 3 illustrates a positive deviation in the $R_0$, $P_0$, and $t_{sc,0}$ results in a redshift of the resonance wavelength. As increasing any of these parameters effectively enlarges the optical path length inside the cavity, thereby shifting the resonant mode towards longer wavelengths. In contrast, a positive deviation in the $E_0$ causes a blueshift. This is because a wide etch reduces the effective refractive index of the patterned region, making the overall structure optically smaller and shifting the resonance to shorter wavelengths. Although in certain cases the perturbed system exhibits a higher Purcell factor, the optimization was specifically designed to maximize the collection efficiency at the target wavelength. Consequently, any deviation from the optimized parameters leads to a slight reduction in efficiency. However, the weights in the cost function can be adjusted to prioritize different performance metrics, depending on experimental requirements. In addition, the efficiency spectrum of the 1550 nm cavity design exhibits a particularly pronounced dip on the longer-wavelength side of the resonance. However, the top transmission monitor indicates an unphysical downward transmission, which is not feasible within our device geometry. We therefore attribute this dip to a numerical artifact arising from the simulation.

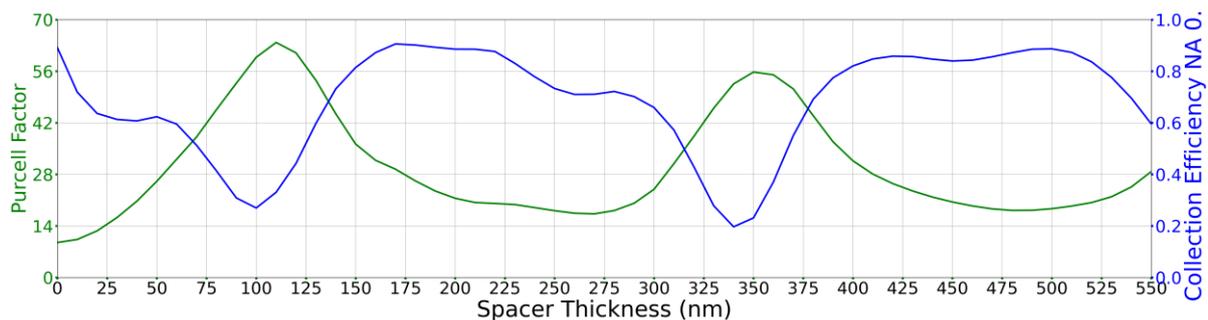

Figure 4. Illustrates the variation of Purcell factor (green) and collection efficiency within $NA = 0.7$ (blue) with $Al_2O_3$ thickness for the optimized Bragg grating for 780nm. The oscillatory trends arise from thickness-dependent optical interference effects in the structure.

While all optimizations were carried out with the objective of maximizing collection efficiency within $NA = 0.7$, it is always possible to achieve a higher Purcell factor by sacrificing some efficiency. Figure 4 demonstrates this trade-off in the 780 nm design by varying $t_{sc}$. Although the Purcell factor peaks around 110 nm, the maximum collection efficiency occurs at an optical distance of approximately $n \cdot \lambda$, corresponding to about 170 nm ($2\lambda$) of $Al_2O_3$, due to constructive interference between the upward-emitted and reflected fields. Therefore, the separation between the semiconductor and the gold mirror can be used for fine-tuning the device. If a higher Purcell factor is desired and lower extraction efficiency is acceptable, one can selected ta different $t_{sc}$ accordingly . In regions where the Purcell factor is high, the upward and reflected fields interfere destructively, leading to a higher cavity Q as light is not efficiently emitted from the cavity.

Beside from reducing collection efficiency, this also result in non-Gaussian farfield modes, which is why we do not consider on these points optimal for overall device performance.

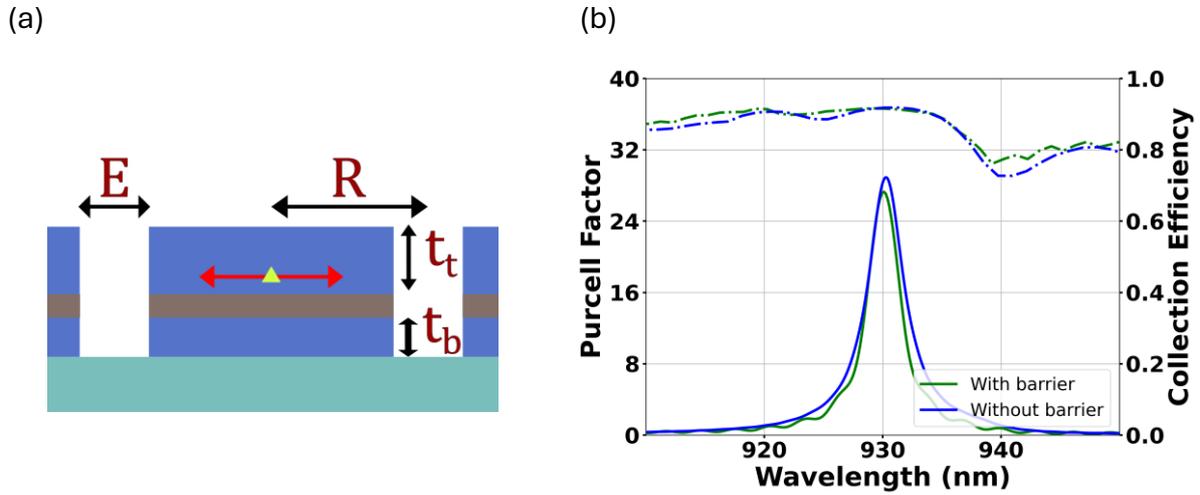

Figure 5. Illustration of 930nm design with 18 nm $Al_{0.8}Ga_{0.2}As$ barrier layer. (a) Cross-section schematic of the design with an additional tunnel barrier, where $t_t$ and $t_b$ denote the top and bottom semiconductor layer thicknesses, respectively. (b) Purcell factor and collection efficiency within $NA = 0.7$ of the 930 nm optimized design with and without the barrier. The results demonstrate that introducing the barrier has minimal impact on the optical properties, and equivalent performance can be achieved after re-optimization.

To investigate the effect of introducing a tunnel barrier layer between the n- and p-doped GaAs regions, an 18 nm $Al_{0.8}Ga_{0.2}As$ layer was added, and the quantum dot was positioned 9 nm above it to ensure compatibility with fabrication constraints, as illustrated in Figure 5a. Upon re-optimizing the design to accommodate the inclusion of the $Al_{0.8}Ga_{0.2}As$ barrier layer, the resulting optimal structural parameters were determined to be an inner radius of 750 nm, a grating period of 370 nm, an etch width of 75 nm, a top semiconductor thickness ($t_t$) of 83 nm, an bottom semiconductor thickness ($t_b$) of 50 nm, and a spacer thickness of 200 nm. Figure 5b compares the simulated Purcell factor and collection efficiency at a NA of 0.7 for 930nm devices with and without the barrier layer, revealing that both configurations yield comparable optical performance.

Although a basic optimization using a fixed grating period and etch width has been implemented, the current design remains far from a global optimum. Recent advances in the design of CBGs have demonstrated that near-perfect Gaussian far-field emission profiles, as well as high Purcell factors and collection efficiencies, can be achieved by individually optimizing each grating period and etch width along the radial direction [32, 33]. Optimizing these parameter independently significantly enhances the design space and the optimization was achieved by solving the problem in cylindrical coordinates as a two dimensional simulation, reducing the computational demands. However, as our design includes bridges, a 2D simulation is not feasible and full 3D-FDTD calculations were required. Assuming a comparable transfer of the figure of merit, a labyrinth CBG structure could likewise be refined to reach such performance levels through meticulous individual tuning of all relevant geometric parameters. This highlights the critical importance of precision fabrication and advanced design strategies for realizing the full theoretical potential of these nanophotonic cavities in practical quantum photonic devices.

# Conclusion

We performed finite-difference time-domain (FDTD) simulations of electrically contacted, labyrinth circular Bragg grating (CBG) cavities designed for resonance wavelengths of 780 nm, 930 nm, and 1550 nm. The goal of this optimization was to maximize the collection efficiency within a numerical aperture (NA) of 0.7 while maintaining a relatively high Purcell factor to enhance light–matter interaction. The simulation results demonstrate the promising potential of this labyrinth-like CBG geometry for quantum dot emitters. Specifically, we achieve collection efficiencies of 90%, 91.2%, and 91.6% for the 780 nm, 930 nm, and 1550 nm devices, respectively, within NA = 0.7.

With the CBG being one of the most extensively studied and widely fabricated nanophotonic cavity designs, its fabrication process has been thoroughly documented and refined over the years. The depth of existing fabrication expertise enables the precise and reliable realization of electrically contacted, bridged CBG structures, making them easier to produce than many alternative cavity geometries. This combination of well-established fabrication protocols and the demonstrated high collection efficiency gives the electrically contacted bridged circular Bragg grating a clear advantage over other designs, making it a highly promising candidate for scalable, high-performance quantum light sources.

## Acknowledgement

We are grateful for financial support from the Federal Ministry of Research, Technology and Space (BMFTR) through the projects QECS (FKZ: 13N16272), PhotonQ (FKZ: 13N15759), and QR.N (FKZ: 16KIS2209). We also acknowledge financial support from the German Research Foundation (DFG) under the project reference DIP FI947/6−1.

We would like to express our gratitude to Matthias Sauter for valuable discussions on software development and optimization pipelines, and Dr. Giora Peniakov for his insightful perspectives on quantum dots in circular Bragg grating (CBG) devices, including thoughtful evaluations of the advantages and limitations of several new design concepts.